\documentclass{amsart}
\def\l{\ell}

\def\ep{\epsilon}
\def\a{\alpha}

\def\be{\begin{equation}}
\def\ee{\end{equation}}
\def\bea{\begin{eqnarray}}
\def\eea{\end{eqnarray}}
\def\ba{\begin{array}}
\def\ea{\end{array}}

\def\bc{{\bf c}}

\def\R{{\mathbb R}}

\def\C{{\mathbb C}}

\newtheorem{theorem}{Theorem}[section]
\newtheorem{lemma}[theorem]{Lemma}

\theoremstyle{definition}

\theoremstyle{remark}

\numberwithin{equation}{section}

\newcommand{\abs}[1]{\lvert#1\rvert}

\newfont{\Bb}{msbm8 scaled\magstep{1}}

\usepackage{graphics}

\begin{document}

\title[MRC Method for Obstacle Scattering Problems]
{Modified Rayleigh Conjecture Method for Multidimensional
 Obstacle Scattering Problems
}

\author{Semion Gutman}
\address{Department of Mathematics\\ University of Oklahoma\\ Norman,
 OK 73019, USA}
\email{sgutman@ou.edu}

\author{Alexander G. RAMM}
\address{ Department of Mathematics\\ Kansas State University\\
Manhattan, Kansas 66506-2602, USA}
\email{ramm@math.ksu.edu}



\begin{abstract} 
 The Rayleigh conjecture on the representation of the scattered 
field in the exterior of an obstacle $D$ is widely used in applications.
However this conjecture is false for some obstacles. 
AGR introduced the Modified Rayleigh
Conjecture (MRC), and in this paper we present successful numerical
algorithms based on the MRC for various 2D and 3D obstacle scattering problems.
The 3D obstacles include a cube and an ellipsoid. The MRC method is easy
to implement for both simple and complex geometries. It is shown to be a
viable alternative for other obstacle scattering methods.

\end{abstract}

\thanks{
}
\maketitle


\section{Introduction.}

In this paper we present a novel numerical method for Direct
Obstacle
 Scattering Problems based on the Modified Rayleigh
Conjecture (MRC). The basic theoretical foundation of the
method was developed in \cite{r430}. The MRC has the appeal
of an easy implementation for obstacles of complicated
geometry, e.g. having edges and corners. In our numerical
experiments the method has shown itself to be a competitive
alternative to the BIEM (boundary integral equations
method), see \cite{r437}. Also, unlike the BIEM, one can
apply the algorithm to different obstacles with very little
additional effort.
 
However, a straightforward numerical implementation of the 
MRC, may be inefficient or fail.
 Nevertheless, one can make the numerical implementation of
the method based on MRC work by introducing some new ideas
within the same theoretical framework. In this paper we
describe such an idea (Random Multi-point MRC) that made
possible, for the first time, to  apply the
method  successfully to 3D obstacles including a sphere and 
a cube.

 In our previous paper \cite{r437} we described another
implementation of the MRC.  That method (Multi-point MRC)
could be used for 2D obstacles of a relatively simple
geometry, but it failed for some 2D obstacles, and it was
not successful for 3D problems because of the problem size limitations.
 In this paper, in addition
to treating 3D problems, we review our
 earlier results and show that they can be significantly
improved by using the new idea for the implementaion of the 
MRC algorithm.

We formulate the obstacle scattering problem in a 3D setting
with the
 Dirichlet boundary condition, but the method 
discussed can
also be used for the Neumann and Robin boundary conditions.

Consider a bounded domain $D \subset \R^3$, with a boundary
$S$ which is assumed to be Lipschitz continuous. Denote the
exterior domain by $D^\prime = \R^3 \backslash D$. Let
 $\alpha, \alpha^\prime\in S^2$ be unit vectors,
where $ S^2$ is the unit sphere in $\R^3$.

The acoustic wave scattering problem by a soft obstacle $D$ 
consists in
finding the (unique) solution to the problem (1.1)-(1.2):
\be\label{s1_e1}
\left(\nabla^2 + k^2 \right) u=0 \hbox{\ in\ } D^\prime, \quad
  u = 0 \hbox{\ on\ } S, 
\end{equation}

\be\label{s1_e2}
u=u_0 + A(\alpha^\prime, \alpha) \frac{e^{ikr}}{r} 
+ o
  \left(\frac{1}{r} \right), \quad r:=|x| \to \infty, \quad
  \alpha^\prime := \frac{x}{r}.
\end{equation}
Here $u_0:=e^{ik \alpha \cdot x}$ is the incident field,
$v:=u-u_0$ is the scattered field,
$A(\alpha^\prime, \alpha)$ is called the scattering amplitude, its
k-dependence is not shown, $k>0$ is the wavenumber. Denote

\be\label{s1_e3}
A_\l (\alpha) := \int_{S^2} A(\alpha^\prime, \alpha)
  \overline{Y_\l (\alpha^\prime)} d\alpha^\prime, 
\end{equation}
where $Y_\l (\alpha)$ are the orthonormal spherical harmonics,
$Y_\l = Y_{\l m}, -\l \leq m \leq \l$. 

Informally, the Random Multi-point MRC algorithm can be
described as follows.

Fix a $J>0$. Let $x_j, j=1,2,...,J$ be a batch of points randomly chosen  inside the
obstacle $D$. For $x\in D^\prime$, let
\begin{equation}\label{s1_hy}
\alpha^\prime =
  \frac{x-x_j}{|x-x_j|},\quad \psi_\l(x,x_j)= Y_\l
(\alpha^\prime) h_\l (k|x-x_j|),  \end{equation} 
where $h_\l (r)$ are the spherical
Hankel functions, normalized so that
$h_\l (r) \sim \frac{e^{ikr}}{r}$ as $r \to +\infty$.

Let
$g(s)=u_0(s),\; s\in S$, and minimize the discrepancy
\be\label{s1_phi}
 \Phi(\bc)=\|g(s)+\sum_{j=1}^J\sum_{\l=0}^L
c_{\l,j}\psi_\l(s,x_j)\|_{L^2(S)}, 
\end{equation}
 over
$\bc\in \C^N$, where $\bc=\{c_{\l,j}\}$. That is, the total
field $u(s)=g(s)+v(s)$ is desired to be as close to zero as
possible at the boundary $S$, to satisfy the required
condition for soft scattering. If the resulting residual
$r^{min}=\min\Phi$ is smaller than the prescribed tolerance
$\ep$, then the procedure is finished, and the sought
scattered field is
$$
v_{\epsilon}(x)=
\sum_{j=1}^J\sum_{\l=0}^{L}c_{\l,j}\psi_\l(x,x_j),\quad
x\in D',
$$
(see Lemma 2.2 below, and the Remark following it).

If, on the other hand, the residual $r^{min}>\ep$, then we
continue by trying to improve on the already obtained fit in
(\ref{s1_phi}). Adjust the field on the boundary by letting
$g(s):=g(s)+v_{\epsilon}(s),\; s\in S$. Create another batch
of $J$ points randomly chosen in the interior of $D$, and
minimize (\ref{s1_phi}) with this new $g(s)$. 
Continue with
the iterations until the required tolerance $\ep$ on the
boundary $S$ is attained. In each iteration accumulate
new interior points $x_j$ and the corresponding
best fit coefficients $c_{\l,j}$. After the desired tolerance is
reached, the sought scattered field $v_\ep$ is computed anywhere in $D'$.

Here is the precise description of the algorithm.

{\bf Random Multi-point MRC.}

For $x_j\in D$, and $\l\geq 0$ functions $\psi_\l(x,x_j)$
are defined as in (\ref{s1_hy}).

\begin{enumerate} \item\label{init4} {\bf Initialization.}
Fix $\ep>0, \; L\geq 0,\;J>0,\; N_{max}>0$. Let $n=0$,
 and $g(s)=u_0(s), \;s\in S$.

\item\label{iter4} {\bf Iteration.}

\begin{enumerate}
 \item Let $n:=n+1$. Randomly choose $J$ points $x_j^{(n)}\in D,\; j=1,2,\dots, J$.

 \item
Minimize
\[
\Phi(\bc)=\|g(s)+\sum_{j=1}^J\sum_{\l=0}^L c_{\l,j}\psi_\l(s,x_j^{(n)})\|_{L^2(S)}
\]
over $\bc\in \C^N$, where $\bc=\{c_{\l,j}\}$.

Let the minimum of $\Phi$ be attained at
$\bc^{(n)}=\{c_{\l,j}^{(n)})\},\; j=1,2,\dots,J$, and 
the minimal value of $\Phi$ be $r^{min}$.

\end{enumerate}

\item {\bf Stopping criterion.}
\begin{enumerate}
 \item If $r^{min}\leq\epsilon$, then  stop.
 Compute the approximate scattered field anywhere in $D'$ by
\be\label{s1_vep}
v_{\epsilon}(x):=\sum_{k=1}^n \sum_{j=1}^J\sum_{\l=0}^{L}c_{\l,j}^{(k)}\psi_\l(x,x_j^{(k)}),\quad x\in D'.
\end{equation}

 \item If $r^{min}>\epsilon$, and $n\not=N_{max}$,
let
\[
g(s):=g(s)+\sum_{j=1}^J\sum_{\l=0}^{L}c_{\l,j}^{(n)}\psi_\l(s,x_j^{(n)}),\quad x\in S
\]
and repeat the iterative
step (\ref{iter4}).
\item If $r^{min}>\ep$, and $n=N_{max}$, then the procedure failed.

\end{enumerate}
\end{enumerate}

\section{Direct scattering problems and the Rayleigh conjecture.}

Let a ball
$B_R := \{x : |x| \leq R\}$ contain the obstacle $D$.
In the region $r>R$ the solution to (\ref{s1_e1})-(\ref{s1_e2}) is:
\be\label{ss2_e1}
u(x, \alpha) = e^{ik\alpha \cdot x} + 
\sum^\infty_{\l =0} A_\l (\alpha)\psi_\l, \quad
\psi_\l:= Y_\l (\alpha^\prime) h_\l (kr), \quad r > R,\quad  
\alpha^\prime =
  \frac{x}{r}, 
\end{equation}
where the sum includes the summation with respect to $m$, $-\l \leq m \leq \l$,
and $A_\l (\alpha)$ are defined in (\ref{s1_e3}).

{\it The Rayleigh conjecture (RC) is:  the series (\ref{ss2_e1}) converges up to 
the 
boundary $S$} (originally RC dealt with periodic structures, gratings).
This conjecture is false for many obstacles, but is true for some 
(\cite{baran,millar,rammb1}). For example, if $n=2$ 
and
$D$ is an ellipse, then the series analogous to (\ref{ss2_e1}) converges in the 
region
$r >a$, where $2a$ is the distance between the foci of the ellipse \cite{baran}.
In the engineering literature there are numerical algorithms, based on the
Rayleigh conjecture. 
Our aim is to give a formulation of a {\it Modified Rayleigh Conjecture} 
(MRC)
which holds for any Lipschitz obstacle and can be used in numerical 
solution of the direct and 
inverse scattering problems. We discuss the 
Dirichlet condition but a 
similar argument is applicable to the Neumann boundary 
condition, corresponding to acoustically hard obstacles.

Fix $\epsilon >0$, an arbitrary small number.

\begin{lemma}\label{s1_lm1} 
There exist $L=L(\epsilon)$ and 
$c_\l=c_\l(\epsilon)$
such that
\be\label{ss2_e6} 
||u_0+\sum_{\l=0}^{L(\epsilon)}c_\l(\epsilon)\psi_\l||_{L^2(S)} \leq 
\epsilon.
\end{equation}

If (\ref{ss2_e6}) and the boundary condition (\ref{s1_e1}) hold, then
\be\label{ss2_e7} 
||v_{\epsilon}-v||_{L^2(S)}\leq \epsilon,  \quad 
v_{\epsilon}:=\sum_{\l=0}^{L(\epsilon)}c_\l(\epsilon)\psi_\l,
\end{equation}
where $v$ is the scattered field defined below formula 
(1.2).
\end{lemma}

\begin{lemma}\label{s1_lm2} 
 If (\ref{ss2_e7}) holds then
\be\label{ss2_e8} 
|||v_{\epsilon}-v|||=O(\epsilon)\,, \quad \epsilon \to 0, 
\quad
\end{equation}
where $|||\cdot|||:= 
||\cdot||_{H_{loc}^m(D')}+||\cdot||_{L^2(D'; 
(1+|x|)^{-\gamma})}$, $\gamma >1$, $m>0$ is an arbitrary integer,
$H^m$ is the Sobolev space, $v$ is the scattered field,
$v_{\epsilon}$ is defined in (2.3), and both
$v_{\epsilon}$ and $ v$ in (\ref{ss2_e8}) are
functions defined in $D'$.

In particular, (\ref{ss2_e8}) implies
\be\label{ss2_e9} 
||v_{\epsilon}-v||_{L^2(S_R)}=O(\epsilon)\,, \quad  \epsilon \to 0, 
\end{equation}
where $S_R$ is the sphere centered at the origin with radius $R$.
\end{lemma}
\noindent{\bf Remark.} The proof of Lemma \ref{s1_lm2} in \cite{r430}
shows that estimate \eqref{ss2_e8} for the scattered field $v$ in the exterior $D'$ 
of the obstacle $D$ follows from the boundary estimate
$$||v_{\epsilon}-v||_{L^2(S)}\leq \epsilon$$
for any outgoing solution $v_\ep$ of $\left(\nabla^2 + k^2 \right) v_\ep=0 \hbox{\
in\ } D^\prime$, regardless of how such solution $v_\ep$ was
constructed.

\begin{lemma}\label{s1_lm3} 
 One has:
\be\label{ss2_e10} 
 c_\l(\epsilon) \to A_\l(\alpha) \,,\quad \forall \l, \quad \epsilon \to 0.
\end{equation}
\end{lemma}

The modified Rayleigh conjecture (MRC) is formulated as a theorem, which
follows from the above three lemmas:

\begin{theorem}\label{s1_thm1} 
 For an arbitrary small $\epsilon>0$ there 
exist
$L(\epsilon)$ and $c_\l(\epsilon),\,\, 0\leq \l \leq L(\epsilon)$,
such that (\ref{ss2_e6}), (\ref{ss2_e8}) and (\ref{ss2_e10}) hold.
\end{theorem}

See \cite{r430} for a proof of the above statements.

The difference between RC and MRC is: (\ref{ss2_e7}) does
not hold if one replaces $v_\epsilon$ by $\sum_{\l=0}^L
A_\l(\alpha)\psi_\l$, and lets $L\to \infty$ (instead of
letting $\epsilon \to 0$). Indeed, the series
$\sum_{\l=0}^\infty A_\l(\alpha)\psi_\l$ diverges at some
points of the boundary for many obstacles. Note also that
the coefficients in (\ref{ss2_e7}) depend on $\epsilon$, so
(\ref{ss2_e7}) is {\it not} a partial sum of a series.

For the Neumann boundary condition one minimizes
$$ \left\|\frac {\partial [u_0+\sum_{\l=0}^{L}c_\l\psi_\l]}{\partial 
N}\right\|_{L^2(S)}$$
with respect to $c_\l$. Analogs of Lemmas 2.1-2.3 are valid and their 
proofs are essentially the same.

See \cite{r461} for an extension of these results to scattering by periodic
structures.

\section{Numerical Experiments.}

In this section we describe numerical results obtained by
the Random Multi-point MRC method for 2D and 3D obstacles.
We also compare the 2D results
 to the ones obtained by our earlier method introduced in \cite{r437}. The
method that we used previously can be described as a Multi-point MRC. It
was not an iterative method, and its applicability was limited by the
problem size, i.e. the number $J$ of the interior points $x_j$.
 It required extensive computational resources (run time and memory)
needed for the minimization part of the algorithm even for a moderate
number $J$.

The Multi-point MRC method is just the first iteration of the
Random method. It can be run efficiently with a relatively modest $J$,
where $J$ is kept constant across iterations. Also, in the Multi-point MRC 
method the
 interior points $x_j,\; j=1,2,...,J,$ were chosen deterministically  
 by an {\it ad hoc} method
 according to the geometry of the obstacle $D$. 

The Random Multi-point MRC eliminates the need for this special
procedure. See the next section for an additional discussion of this
issue.

As we mentioned
previously, \cite{r437} contains a favorable comparison of
the Multi-point MRC method with the Boundary Integral Equation Method,  despite 
the fact that the numerical implementation of the MRC 
method in \cite{r437} is considerably less efficient than the one
presented in this paper. 

A numerical implementation of the Random Multi-point MRC
method follows the same outline as for the Multi-point MRC,
which was described in \cite{r437}. Of course, in a 2D case,
instead of (\ref{s1_hy}) one has

\[
\psi_l(x,x_j)=H_l^{(1)}(k\abs{x-x_j})e^{il\theta_j},
\]
where $(x-x_j)/\abs{x-x_j}=e^{i\theta_j}$.

For a  numerical implementation choose
$M$ nodes $\{t_m\}$ on the surface $S$ of the obstacle $D$. 
After the interior points 
$x_j,\; j=1,2,...,J$ are chosen, form   $N$ vectors 
\[
{\bf a}^{(n)}=\{\psi_l(t_m,x_j)\}_{m=1}^M,
\]
$n=1,2,\dots,N$ of length $M$. Note that $N=(2L+1)J$
for a 2D case, and $N=(L+1)^2J$ for a 3D case.
It is convenient to normalize the norm in $\R^M$ by
\[
\|{\bf b}\|^2=\frac 1M \sum_{m=1}^M|b_m|^2,\quad {\bf
b}=(b_1,b_2,...,b_M).
\]
Then $\|u_0\|=1$.

Now let ${\bf b}=\{g(t_m)\}_{m=1}^M$, in
the Random Multi-point MRC (see section 1), and minimize
\be\label{s3_minm}
\Phi({\bf c})=\|{\bf b}+A{\bf c}\|,
\end{equation}
for ${\bf c}\in \C^N$, where $A$ is the matrix containing vectors ${\bf a}^{(n)},\;
n=1,2,\dots,N$ as its columns.

We used the Singular Value Decomposition (SVD) method (see e.g.
\cite{numrec}) to minimize (\ref{s3_minm}).  Small singular values $s_n<w_{min}$ of the
matrix $A$ are used to identify and delete linearly dependent or almost
linearly dependent combinations of vectors ${\bf a}^{(n)}$. This spectral
cut-off makes the minimization process stable, see the details in
\cite{r437}. 

 Let $r^{min}$ be the residual, i.e. the minimal value 
of
$\Phi({\bf c})$ attained after $N_{max}$ iterations of the
Random Multi-point MRC method (or when it is stopped). For a
comparison, let $r^{min}_{old}$ be the residual obtained in
\cite{r437} by an earlier method.

We conducted 2D numerical experiments for four obstacles: two ellipses of
different eccentricity, a kite, and a triangle. 
The M=720 nodes $t_m$ were uniformly distributed on the interval
$[0,2\pi]$, used to parametrize the boundary $S$.
Each case was tested for
wave numbers $k=1.0$ and $ k=5.0$. Each obstacle was subjected to incident
waves corresponding to $\a=(1.0,0.0)$ and $\a=(0.0,1.0)$. 

The results for the Random Multi-point MRC with $J=1$ are shown in Table 1, in the
last column $r^{min}$. In every experiment the target
residual $\ep=0.0001$ was obtained in under 6000 iterations,
 in about 2 minutes run time on a 2.8 MHz PC.

In \cite{r437}, we conducted numerical experiments for the same four 2D
obstacles by a Multi-point MRC, as described in the beginning of this
section. The interior points $x_j$ were chosen differently in
each experiment. Their choice is indicated in the description of each 2D
experiment.
The column $J$ shows the number of these interior
points. Values $L=5$ and $M=720$ were used in all the
experiments. These results are
shown in Table 1, column  $r^{min}_{old}$.

 Thus, the Random Multi-point MRC method achieved a significant
improvement over the earlier Multi-point MRC. 

\begin{table} 
\caption{Normalized residuals attained in the
numerical experiments for 2D obstacles,
 $\|{\bf u_0}\|=1$.}

\begin{tabular}{c  r  r  c  r   r}

\hline
Experiment & $J$ & $k$ & $\a$ & $r^{min}_{old}$ & $r^{min}$ \\

\hline
I & 4 & 1.0 & $(1.0,0.0)$ & 0.000201  & 0.0001\\
  & 4 & 1.0 & $(0.0,1.0)$ & 0.000357  & 0.0001\\
  & 4 & 5.0 & $(1.0,0.0)$ & 0.001309  & 0.0001\\
  & 4 & 5.0 & $(0.0,1.0)$ & 0.007228  & 0.0001\\
\hline
II & 16 & 1.0 & $(1.0,0.0)$ & 0.003555  & 0.0001\\
  & 16 & 1.0 & $(0.0,1.0)$ & 0.002169  & 0.0001\\
  & 16 & 5.0 & $(1.0,0.0)$ & 0.009673  & 0.0001\\
  & 16 & 5.0 & $(0.0,1.0)$ & 0.007291  & 0.0001\\
\hline
III & 16 & 1.0 & $(1.0,0.0)$ & 0.008281  & 0.0001\\
    & 16 & 1.0 & $(0.0,1.0)$ & 0.007523  & 0.0001\\
    & 16 & 5.0 & $(1.0,0.0)$ & 0.021571  & 0.0001\\
    & 16 & 5.0 & $(0.0,1.0)$ & 0.024360  & 0.0001\\
\hline
IV & 32 & 1.0 & $(1.0,0.0)$ & 0.006610  & 0.0001\\
   & 32 & 1.0 & $(0.0,1.0)$ & 0.006785  & 0.0001\\
   & 32 & 5.0 & $(1.0,0.0)$ & 0.034027  & 0.0001\\
   & 32 & 5.0 & $(0.0,1.0)$ & 0.040129  & 0.0001\\
\hline
\end{tabular}

\end{table}

{\bf Experiment 2D-I.} The boundary $S$ is an ellipse described by
\be
{\bf r}(t)=(2.0\cos t,\ \sin t),\quad 0\leq t<2\pi\,.
\end{equation}
The Multi-point MRC used $J=4$ interior points 
$x_j=0.7{\bf r}(\frac{\pi(j-1)}2),\; j=1,\dots,4$.
The run time was 2 seconds.

{\bf Experiment 2D-II.} The kite-shaped boundary $S$ (see \cite{coltonkress}, Section 3.5) is  
described by
\be
{\bf r}(t)=(-0.65+\cos t+0.65\cos 2t,\ 1.5\sin t),\quad 0\leq t<2\pi\,.
\end{equation}    
The Multi-point MRC used $J=16$ interior points $x_j=0.9{\bf r}(\frac{\pi(j-1)}8),\; j=1,\dots,16$.
The run time  was 33 seconds.


{\bf Experiment 2D-III.} The boundary $S$ is the triangle
with vertices at $(-1.0,0.0)$ and $(1.0,\pm 1.0)$.
The Multi-point MRC used the interior points $x_j=0.9{\bf r}(\frac{\pi(j-1)}8)$,
$ j=1,\dots,16$.
The run time  was about 30 seconds.

{\bf Experiment 2D-IV.} The boundary $S$ is an ellipse described by
\be
{\bf r}(t)=(0.1\cos t,\ \sin t),\quad 0\leq t<2\pi\,.
\end{equation}
The Multi-point MRC used $J=32$ interior points 
$x_j=0.95{\bf r}(\frac{\pi(j-1)}{16}),\; j=1,\dots,32$.
The run time was about 140 seconds.

The 3D numerical experiments were conducted for 3 obstacles: a sphere, a cube, and an ellipsoid.
We used the Random Multi-point MRC with $L=0,\; w_{min}=10^{-12}$, and $J=80$.
 The number $M$ of the points on the boundary $S$ is
indicated in the description of the obstacles. The scattered
field for each obstacle was computed for two incoming
directions $\a_i=(\theta,\phi),\;i=1,2$, where $\phi$ was
the polar angle. The first unit vector $\a_1$ is denoted by
(1) in Table 2, $\a_1=(0.0,\pi/2)$. The second one is
denoted by (2), $\a_2=(\pi/2,\pi/4)$. A typical number of
iterations $N_{iter}$ and the run time on a 2.8 MHz PC are
also shown in Table 2. For example, in experiment I with
$k=5.0$ it took about 700 iterations of the Random
Multi-point MRC method to achieve the target residual
$r^{min}=0.001$ in 7 minutes.

{\bf Experiment 3D-I.} The boundary $S$ is the sphere of radius $1$,
with $M=450$.

{\bf Experiment 3D-II.} The boundary $S$ is the surface of the cube
$[-1,1]^3$ with $M=1350$.

{\bf Experiment 3D-III.} The boundary $S$ is the surface of the
ellipsoid $x^2/16+y^2+z^2=1$ with $M=450$.

\begin{table}
\caption{Normalized residuals attained in the numerical experiments for 3D obstacles,
$\|{\bf u_0}\|=1$.}

\begin{tabular}{c  r  r  l  r   r}

\hline
Experiment & $k$ & $\a_i$ &  $r^{min}$ & $N_{iter}$ & run time \\

\hline
I & 1.0 &  & $0.0002$ & 1   & 1 sec\\
  & 5.0 &  & $0.001$ & 700  & 7 min\\

\hline
II & 1.0 & (1) & $0.001$ & 800   & 16 min\\
   & 1.0 & (2) & $0.001$ & 200  & 4 min\\
   & 5.0 & (1) & $0.0035$ & 2000   & 40 min\\
   & 5.0 & (2) & $0.002$ & 2000  & 40 min\\

\hline
III & 1.0 & (1) & $0.001$ & 3600   & 37 min\\
    & 1.0 & (2) & $0.001$ & 3000   & 31 min\\
    & 5.0 & (1) & $0.0026$ & 5000   & 53 min\\
    & 5.0 & (2) & $0.001$ & 5000   & 53 min\\
\hline
\end{tabular}

\end{table} 
In the last experiment the run time could be reduced by taking a smaller value
for $J$. For example, the choice of $J=8$ reduced the running time to
about 6-10 minutes.

Numerical experiments show that the minimization results depend on the
choice of such parameters as $J,\; w_{min}$, and $L$.

\section{Discussion of the results.}

Let $D$ be an obstacle, $S$ be its boundary, and $u_0$ be the incident field. 
It is proved in \cite{r430} that if
 $v_\ep$ is an outgoing solution of the Helmholtz equation in the exterior 
domain  $D'$ and  
 $u_0 +v_\ep$ approximates zero in $L^2(S)-$norm on the boundary $S$,  
then $v_\ep$   approximates the exact scattered
field $v$ in  $D'$, see Lemma 2.2 and the
Remark after it.
The Modified Rayleigh Conjecture approach to obstacle scattering
problems is based on the following observation: 
the functions $\psi_\l(x,z),\;z\in D$
and their linear combinations are outgoing solutions to
the Helmholtz equation in the exterior domain. Therefore, one
just needs to find a  combination of such functions that gives the best fit to 
$-u_0$ on the boundary $S$. Then this combination 
approximates the scattered field everywhere in the exterior $D'$ of
the obstacle $D$ and the error of this approximation is given in 
Theorem 2.4.

Various implementations of the MRC method provide different algorithms for 
the
construction of such best fit. The original theory of MRC method given in
\cite{r430} guarantees 
that one can  use all $\psi_\l(x,z)$ at $z=0$
to obtain the required fit. However, it is not always possible to 
obtain the desired accuracy
numerically, because this would require a very high accuracy of
computations due to the fact that the Hankel functions differ
from each other by many orders of magnitude if $\l$ is large. Thus, one 
wants to
restrict the order $L$. We found numerically that $L=5$ is
a reasonable value from the numerical point of view.

To keep $L$ reasonably small, we suggested in \cite{r437} to use a batch
of interior points $x_j\in D,\;j=1,2\dots,J$, and the associated
source functions $\psi_\l(x,x_j)$ to find the best fit (Multi-point MRC). 
The results are presented in
\cite{r437}. They improve with the increase in the number $J$ of 
the interior points.
However, for $J>30$ the problem size becomes too large for our computer
system to handle it efficiently, thus limiting  the applicability of the
Multi-point MRC. Still, if one can achieve a satisfactory fit with this
method (2D or 3D), then it is more efficient than BIEM. Another issue in
the Multi-point MRC is the placement of the interior points $x_j$. In  \cite{r437} we
used an {\it ad hoc} procedure, by spreading the points just behind the
boundary $S$ of the obstacle.

In the present paper we made a further advance in the MRC algorithm. Our
Random Multi-point MRC is an iterative method. Thus it allows to keep
the number $J$ of the interior point relatively small, but still achieve
a good boundary fit. Undoubtedly, there are many strategies as to how to
place the interior points $x_j$. Clearly, they should be chosen
distinctly in subsequent iterations. Numerical experiments showed that restricting these points
to a subset of the obstacle $D$ did not produce satisfactory relults.
We tried to place
the points in the interior of $D$ in a random fashion, to assure that
the entire interior is utilized. The results of this algorithm are
presented in this paper. They show a significant improvement in the
obtained fit as compared with our earlier method. Also, we were able to
obtain accurate solutions for 3D problems. Clearly, the random choice of 
the points is
not essential to the algorithm. We think that a suitable deterministic
choice of points could be as successful and are working on finding such a 
choice.

It may happen that subsequent iterations bring only a negligible
improvement to the already obtained minimization values of $\Phi$
in the method we proposed in this paper, and we do not have a 
theoretically justified way around this difficulty.

We are trying to find an optimal placement for the interior points
and we hope to report the results of this research
upon its completion.

\section{Conclusions.}
For a 2D, or 3D obstacle, Rayleigh conjectured that the acoustic field
$u$ in the exterior of the obstacle is given by 
\be
u(x, \alpha) = e^{ik\alpha \cdot x} + 
\sum^\infty_{\l =0} A_\l (\alpha)\psi_\l, \quad
\psi_\l:= Y_\l (\alpha^\prime) h_\l (kr), \quad  
\alpha^\prime =
  \frac{x}{r}.  \end{equation} While this conjecture (RC) is
false for many obstacles, it has been modified to obtain a
representation for the solution of
(\ref{s1_e1})-(\ref{s1_e2}) and to obtain its error.
 This representation 
(Theorem
\ref{s1_thm1}) is called the Modified Rayleigh Conjecture
(MRC). In fact, it is not a conjecture, but a theorem.

We propose here an implementation of the MRC method
which gives an efficient approach to solving obstacle scattering problems
in 3D problems with complicated geometries.
Our implementation of the MRC method worked in the cases considered more 
efficiently than the BIEM method.

The implementation of the MRC method presented in this paper, 
the Random Multi-point
MRC, has been successfully applied to various 2D and 3D
obstacle scattering problems. This algorithm is a
significant improvement over the earlier implementation of MRC method, 
given in \cite{r437}.
The improvement is achieved by allowing the required
minimizations to be done iteratively. 
However, even the earlier implementation
(see \cite{r437})
compared favorably
 to the BIEM.

The Random Multi-point MRC has an additional attractive
feature:  it can easily treat obstacles with complicated
geometry (e.g.  edges and corners). Unlike the BIEM, 
the Random Multi-point MRC
can be relatively easily used for solving  obstacle scattering problems
with complicated geometries and rough boundaries.

Further research on  MRC algorithms is under way.  
The authors hope  that a numerical implementation of the MRC
method will yield a more efficient and economical
algorithm for solving obstacle scattering problems
than the currently used methods.

\section{Acknowledgement.}

 The authors thank a referee for useful remarks.

\end{document}